# Scanning for time:
# Science and art on a photocopier

By Eric Muller
The Exploratorium Teacher Institute
Pier 17
San Francisco, CA 94111

What do you get when you cross a rubber band with a photocopier?
You get a whole series of physics lessons and some great images!

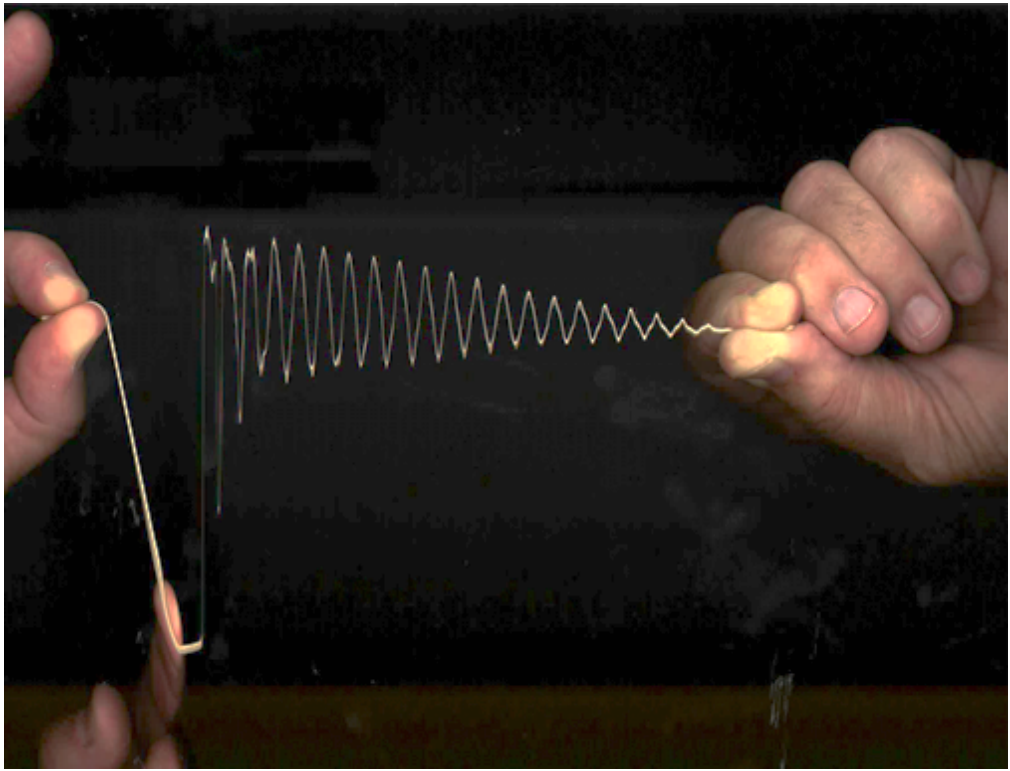

It's easy to get this activity up and running, be amazed, and apply it to any number of scientific concepts.  It lends itself to a whole host of investigations.  This easy to do activity can also be aligned to NGSS outcomes and three-dimensional learning.  It's science, art, fun, and your results may be surprisingly…. well, surprising.

**What you'll need:**
- A thick, long rubber band (try an assortment)
- High-speed photocopier/scanner (found in most offices, schools or copy shops)
    Make sure your copier scans documents or pictures by moving a lighted bar across the scanner bed. If your copier can scan an

image in a second or two, you're in business. If needed, adjust the dpi (dots per inch) to get the machine scanning faster. Some machines scan at different rates for color vs. black/white modes. Scan quality might also differ from manufacturer to manufacturer. (Note : home/office photocopiers might not be fast enough to produce useful results).[1]
- Scissors
- Partner
- Cellphone or digital camera
- Metric ruler
- Optional:
    - Bungee/elastic cord
    - Fidget spinner
    - Dry spaghetti with a gummy bear on top
    - Soda can
    - Guitar or musical instrument (see a "Musical Sidebar" below)

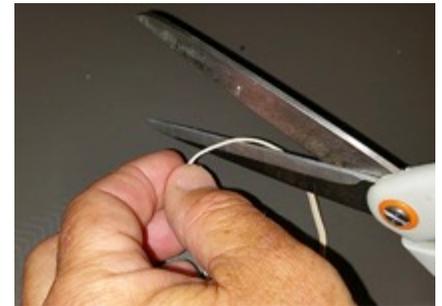

**Here's what to do:**

1. Cut the rubber band with scissors so that it is a linear strand.
2. Have a partner help you with this:
    a. Stretch the rubber band across the high-speed copier bed and perpendicular to the axis of the scan bar (usually held left to right) (Note: The photocopied image on page one, was the result of a loosely stretched rubber band and the next two images show the same rubber and output stretched more tightly.)
    b. The rubber band should be held very close to the glass but not touching it, so that it is free to vibrate.
    c. Pinch and pull the rubber band to one side and be ready to release.

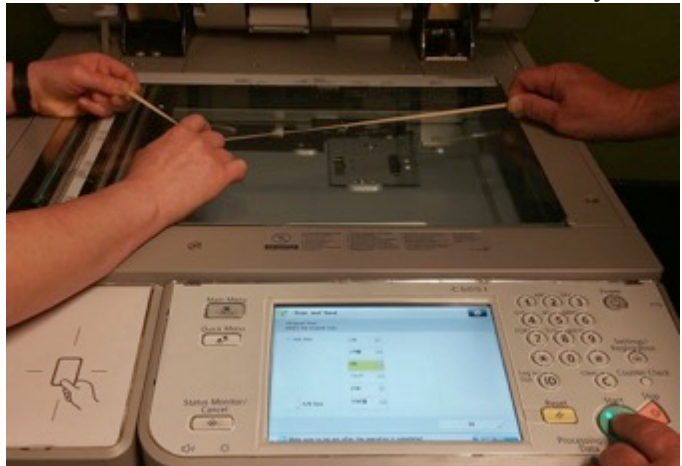

    d. Press the start button on the photocopier.

e. When the lighted scanner bar moves under the rubber band, release the rubber band so that it can vibrate back and forth.
3. As soon as the copier's light turns off, you're done!
4. Remove the copy from the tray and see what was recorded.

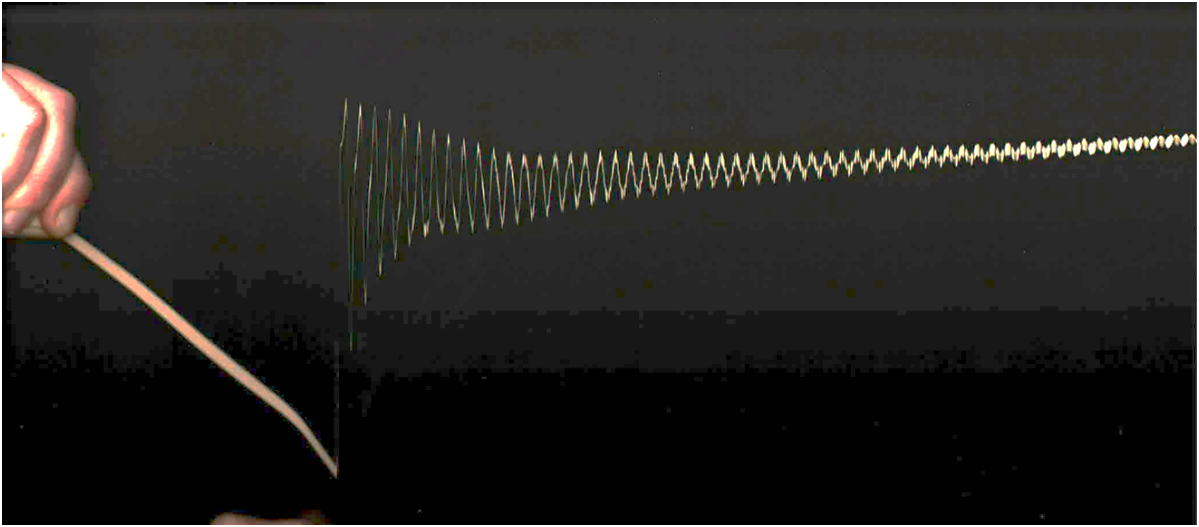

**<u>What's going on?</u>**
Cool, right? …and all from a vibrating rubber band and photocopier!

A photocopier is designed to record stationary flat objects. It doesn't work like your eye/brain system. It doesn't randomly scan a scene for important locations and events and try to put everything in context. A photocopier only records what it "sees" directly above the optical scan bar and that's it.

A photocopier contains a host of electronic and mechanical systems that all need to be in sync. After the start button is pressed, a scanner or scan bar moves at a set and constant rate beneath a glass bed.

When plucked, your rubber band really vibrates back and forth as a fundamental harmonic waveform (1/2 wavelength), just like a guitar string. Pluck your rubber band and prove that to yourself!

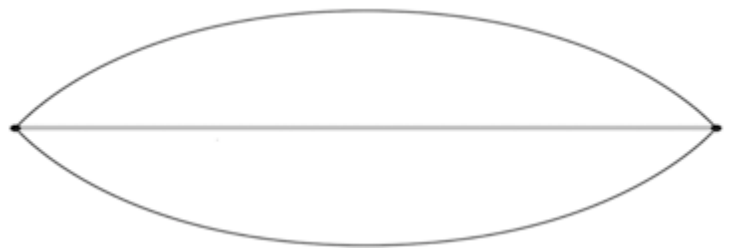

What is recorded happens over the course of a few seconds. But because there are two perpendicular moving parts that combine to make the image (the vibrating rubber band and, 90° degrees to that, the scan bar) you don't see a simple harmonic. What is recorded looks like a sine-wave pattern that's bigger on one end and flattens out to a line on the other.

To the right is a hypothetical series of images you might see if you were able to look up and through the glass scanner bed. Here are ten frames or 2 1/4 cycles of back and forth (or up and down in this view) motion of the rubber band (**pink**). The scan bar moves sideways frame by frame at a constant velocity (**green**) and records the view at each instance (**black dot**). The image, is a continuous recording (**light blue**).

What is produced, is a composite of both motions over time (**dark blue.**)

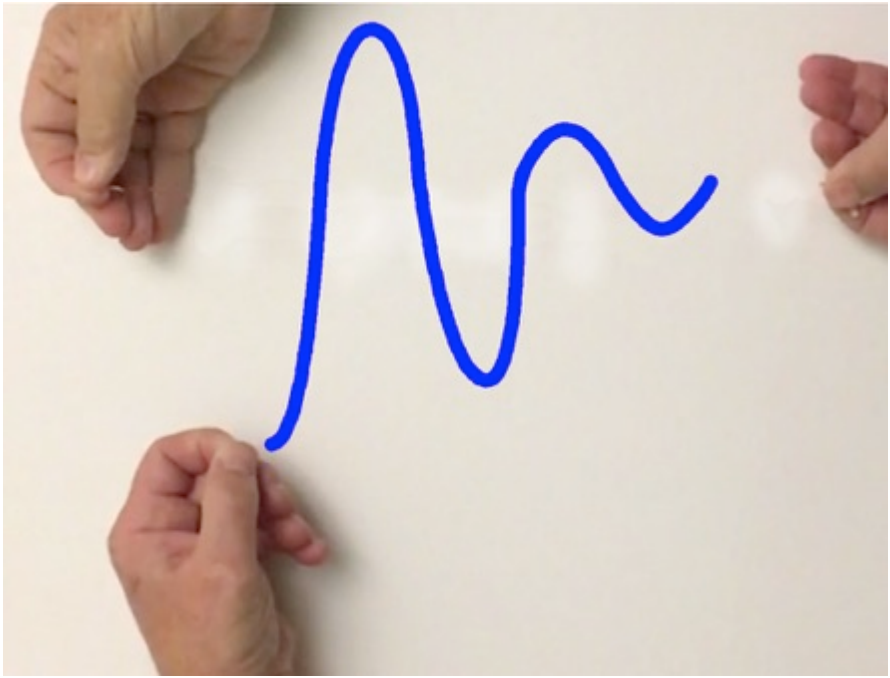
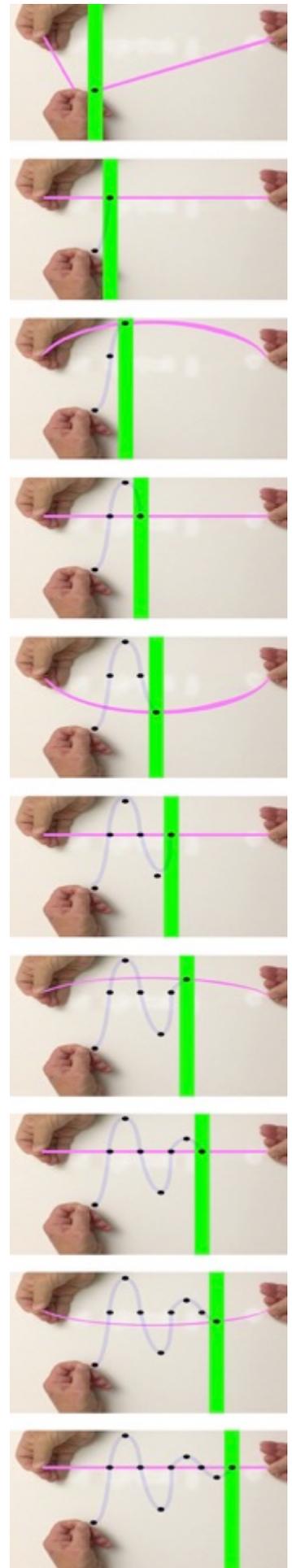

Since the scan bar moves a certain distance in a certain amount of time, the copy you see is a graph; a time and distance graph! Each **wave crest** represents one full **cycle** of back and forth motion of the rubber band, but in a different location or moment in time. Each wave crest is **equidistant**, because the rubber band vibrates at a **constant rate**. You can hear the **frequency or pitch** of the plucked band. If you change the **tension, length or mass (thickness)** of the rubber band, you change the frequency or rate

© 2018: Eric Muller, Exploratorium Teacher Institute- original manuscript
Submitted to the American Association of Physics Teachers journal, August 2018
To be published in The Physics Teacher magazine (extension or journal), Jan, 2019, vol. 57 #1

of back and forth motion and therefore the distance from wave to wave or **wavelength**.

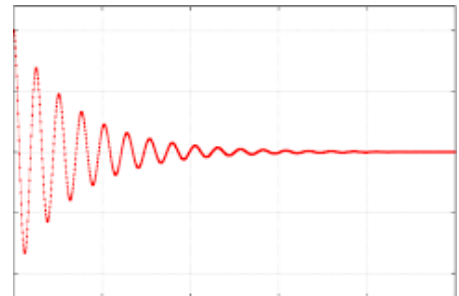

The scan bar is moving from the region where you plucked it (anti-node) to where you're holding it (node) and the oscillation also naturally dampens with time. You can see the combined effect of these two factors by the successive decrease in wave height. If you listen, you may notice that the rubber band does indeed get quieter over a short period of time. The vibration **decays**, and the **amplitude** of each successive wave gets smaller and smaller. All this is seen, because both phenomena, the moving rubber band and the scan bar, take time.

A similar effect happens with digital cameras, like those on cellphones. If you take a picture of a very fast moving object, like a spinning propeller, you'll get a recorded image that looks distorted. Most cellphone cameras scan the scene just like photocopiers do. Although the scan is all electronic (no moving parts) and it happens much faster, it still takes time. And on a digital camera it can happen when taking a still shot or while taking a movie.

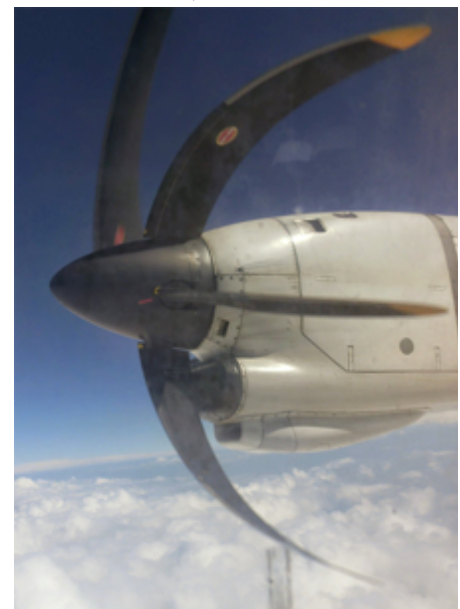

This phenomenon is known as the "Rolling Shutter Effect. [2,3,4]

### Want to do some math?
You can do some interesting calculations with your photocopy, since it's a graph of events that happened over a plate of glass. Here's one to try: find out the frequency at which your rubber band is wiggling.

### *First figure out the speed of your scanner.*
All scanners scan documents at a constant rate, but figuring this speed accurately can be a challenge unto itself.

There are several ways to figure this out, but here's my favorite:

1. Use a cellphone or digital camera to make a video recording of your photocopier in action. Using the standard recording speed of 30 frames[5] per second (fps), record the machine as it scans from one side of the bed to the other. Most beds are either, tabloid size 17.0 inches (432cm) long in the US or Canada or A3 size (420mm or 16.54 inches) in other countries.
2. Play back your video on a computer or video system that lets you step through every individual frame.
   (Note: Make sure your software doesn't "drop" or remove any frames due to digital compression!).

3. Count the number of frames from when your scanner starts to when it stops scanning.
   Here are a couple of things to know to make a better frame count:
   a. Many photocopiers give the scan bar a short distance to "run up" to speed.
   b. Be mindful that determining the first and last frame is made more difficult due to the extreme brightness of the scan bar.

   Enter the number of frames from start to end of scan = ____________

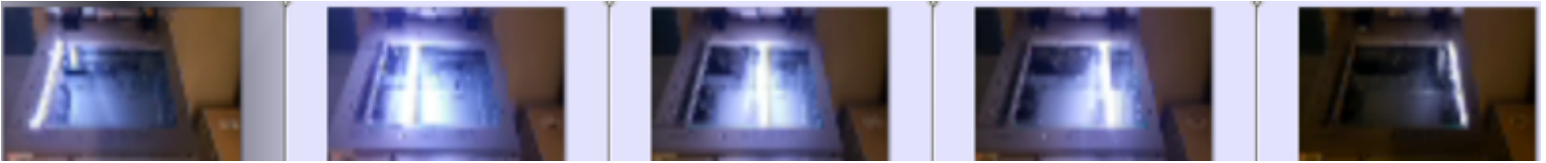

Calculate your photocopiers scan speed:
1. Figure out the length of your scan bed in meters

   Scan bed length = __________ meters

2. Figure out how long it takes to scan a document.
   Since your video is captured at rate of 30 frames per second, then each frame represents 1/30 or .0333 seconds.
   Multiply the number of frames above by .0333 = _________

3. Speed of scanner is distance / time = _____________ m/sec.

> Here's an example from the Exploratorium's office scanner.
> From beginning to end, our video shows that scan took 53 frames to cover the 17.0 inches or .432 meters scan bed.
> 53 frames x .033 seconds= **1.75** seconds.
> The speed of scanner is therefore = .432meters/1.75 seconds
>     Or  V = **.25 meters/second**

### Figure out the wavelength (λ) of your copied waveform.
Photocopies are designed to scan and print documents incredibly close to the original document's dimensions.
(Note: make sure your copier is set to 100% copy ratio).
This means you can directly measure the distance from wave crest to wave crest.
However, it is better to measure the distance of multiple waves and divide by the number of waves. This average will give you better data than measuring an individual wave.
1. Use a metric ruler to measure the distance of multiple waves.

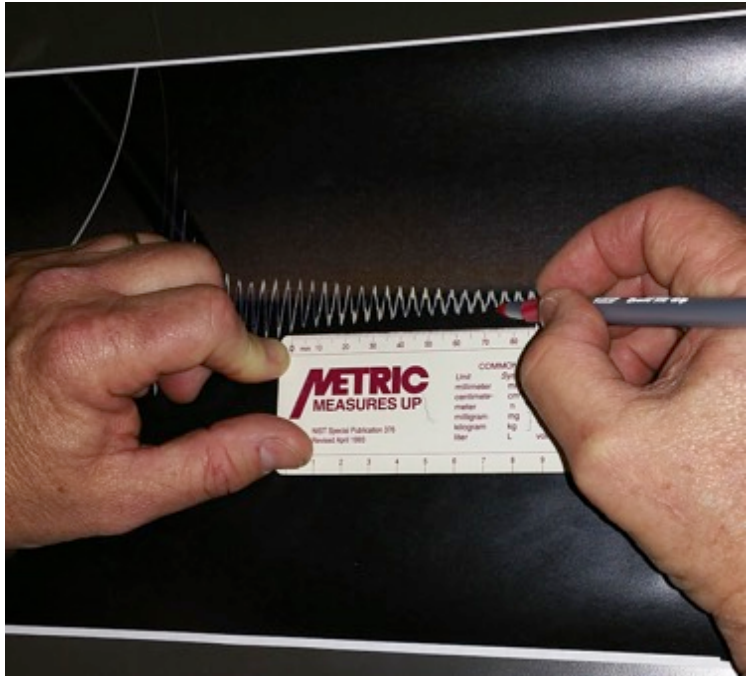

   a. Number of waves= _____________

   b. Distance from first crest to last= ________ meters

2. To get the average wavelength, divide the distance by number of waves.

   Wavelength = _______m

> For our example:
> We measured 15 waves over a distance of 7.0cm or .070meters
> So, the graph has a **wavelength of .070meters/15waves**
> Or
> **λ = .0047 meters/wave**
>
> Note that this is not the wavelength of the standing wave fundamental. The captured waves and associated wavelengths are an artifact of how the scanner records the moving fundamental wave. Each crest to crest (or trough to trough) recorded wave of the vibrating rubber band represents a different time and location of the scan bar as it moves under each cycle of the actual fundamental wave.

Finally, figure out the frequency (**f**) of your vibrating rubber band.
Although the rubber band is really moving at a fundamental harmonic, it's multiple waveform recording was created by a moving a scan bar of known velocity (V) tracing waves of a known wavelength (λ). You can use the wave equation:

$$f = v/\lambda$$

Therefore our example has a frequency of:

**f = .25 meters/second / .0047 meters/wave**
or
**f = 53 waves/second or 53 Hz**

And the best part, this can be confirmed with sound analyzing software!! Here's our set-up of recording and then plucking a rubber band. Note the external microphone to help with recording.

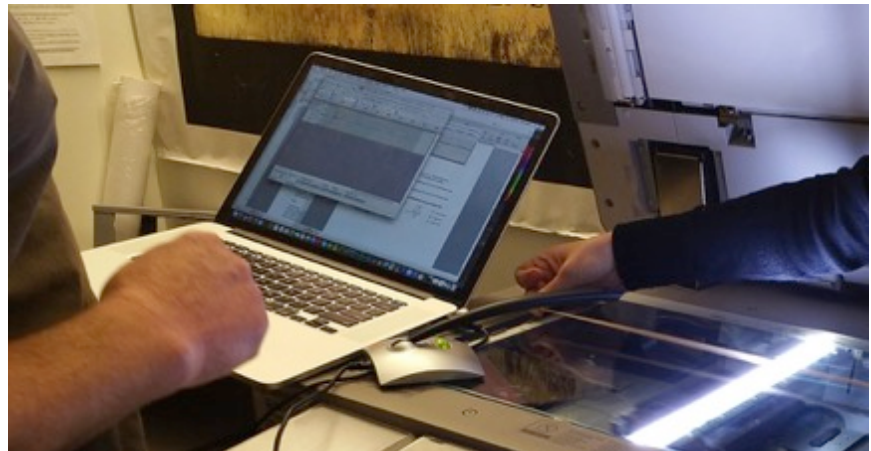

Shown here is Audacity, a free audio analysis program.

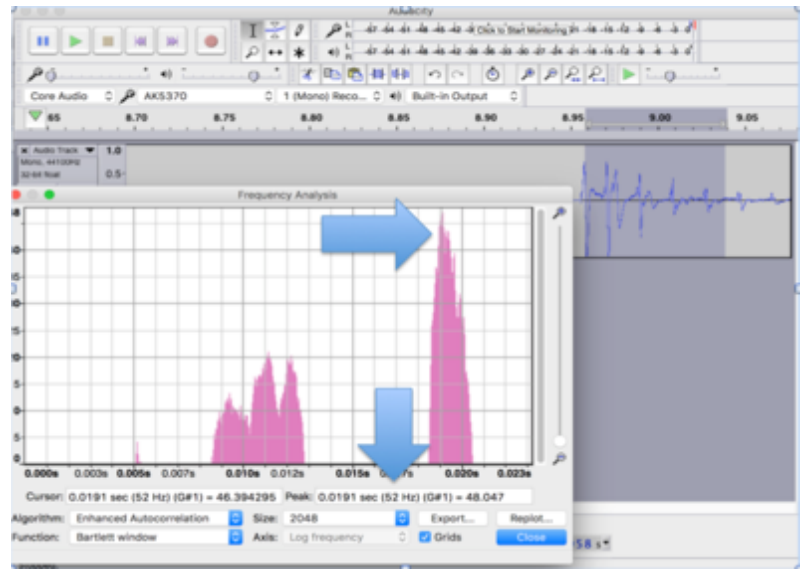

Our calculated frequency was 53hz and our audio sample of the same plucked rubber band was 52hz…pretty good. Considering all the small precision errors and these numbers still agree within an error of less than 2%...that's amazing!

**Going further**

Here are some things to try:
- If you know the frequency of your vibrating rubber band (by using an audio analysis program), you can figure out the scanning speed of other photocopiers.
- Experiment with tension, mass and length of your rubber band. How would it affect your photocopied **outcome**?
- Use this activity to spark a variety of sound and wave investigations and experiments.
- Find things that spin, roll or move. Predict what you will see from the copier. Just by pressing "start," you can confirm your guess and make some great images!

$$f_1 = \frac{\sqrt{\dfrac{T}{m/L}}}{2L}$$

$T$ = string tension
$m$ = string mass
$L$ = string length

Here are some examples:

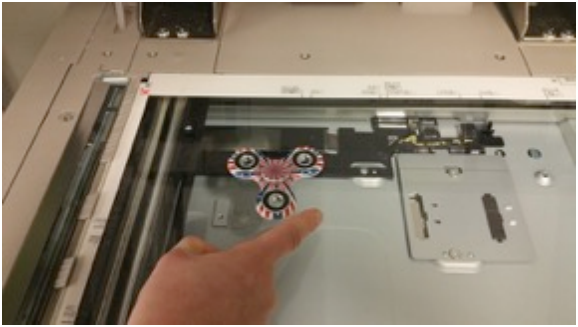

**Spinning a Fidget spinner**

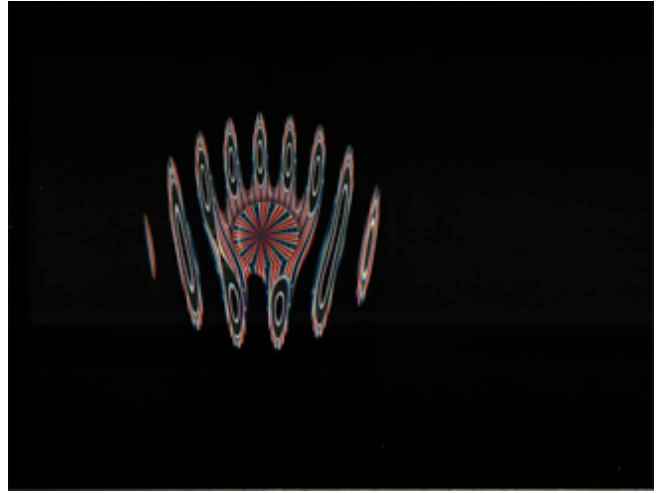

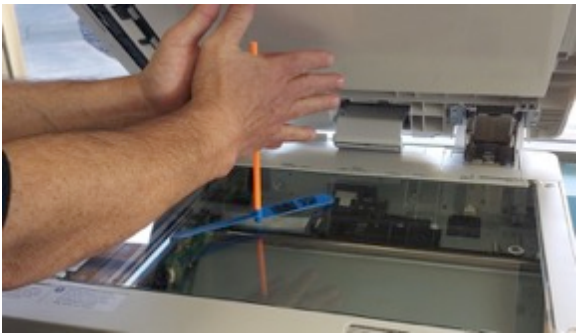

**Spinning a hand-spun helicopter toy upside-down on the scanner bed.**

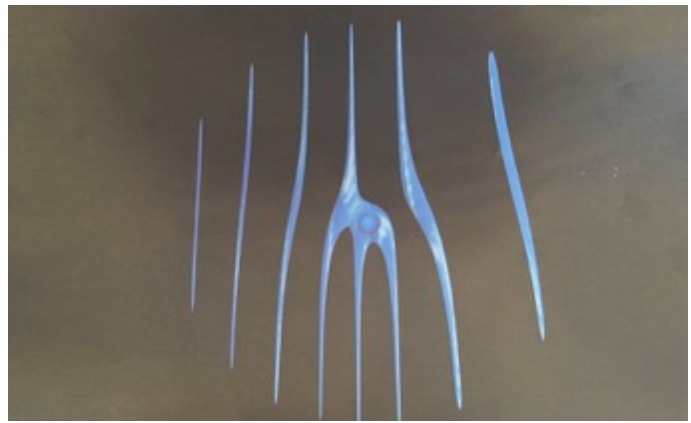

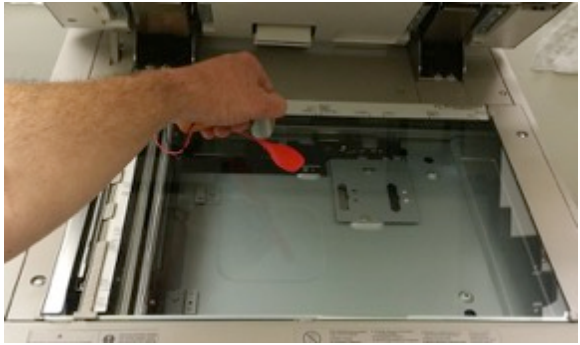

**Spinning a propeller with an small electric motor**

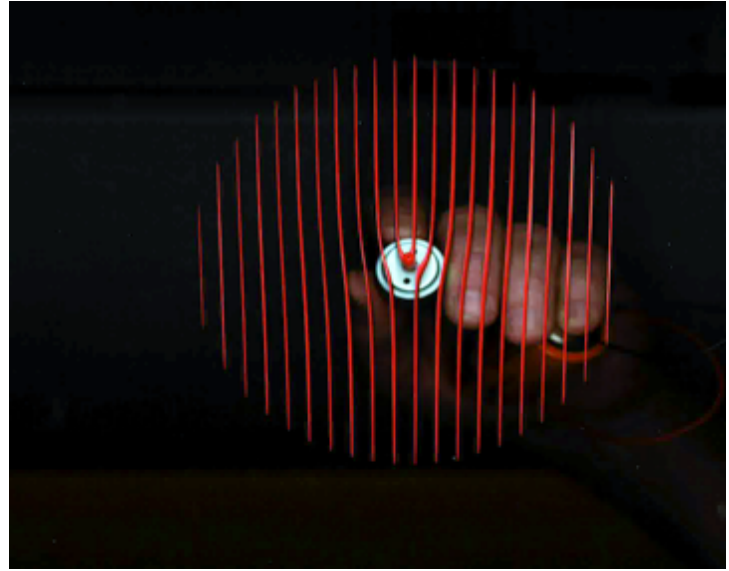

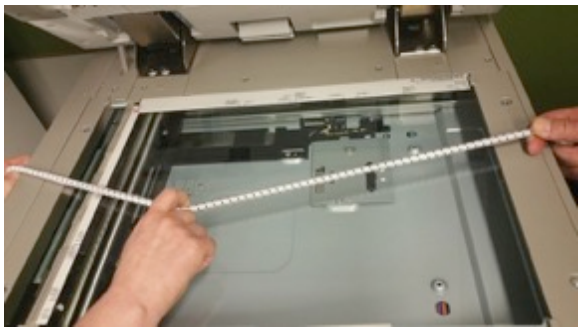

**Plucking a Bungee Cord in different directions**

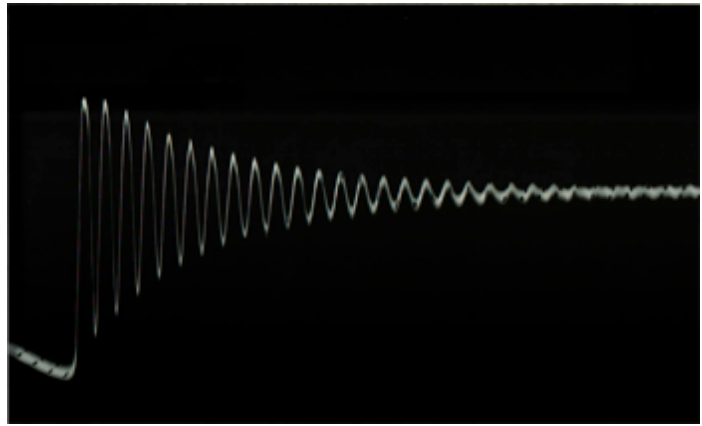

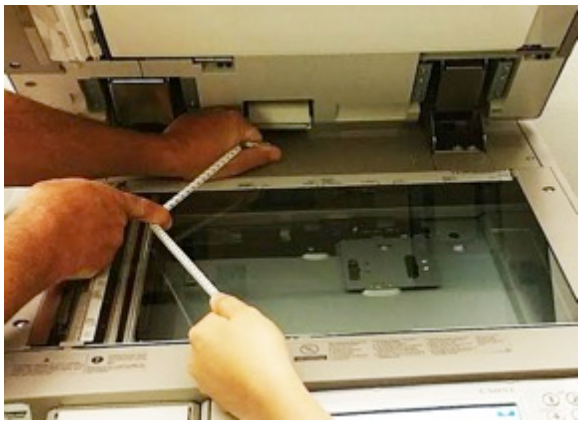

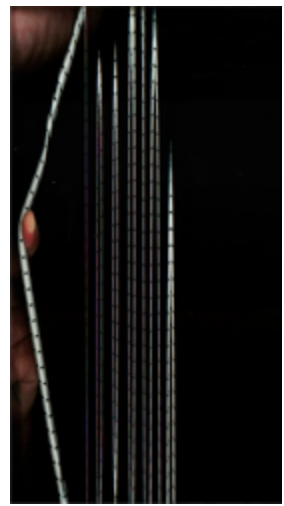

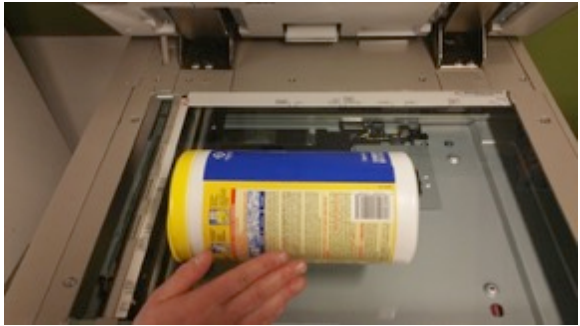

**Rolling a cylinder across a scanner bed.**

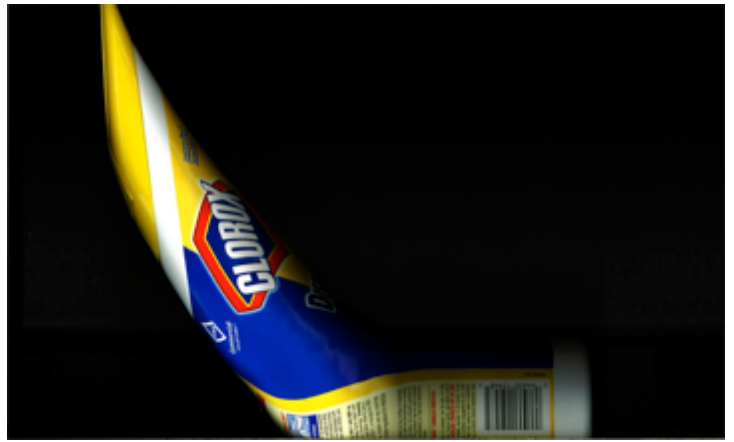

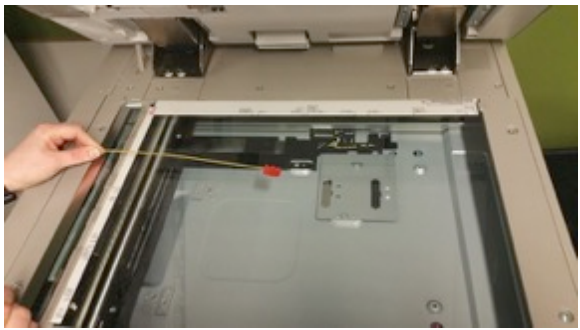

**Wiggling a gummy bear on the end of a piece of spaghetti in different directions.**

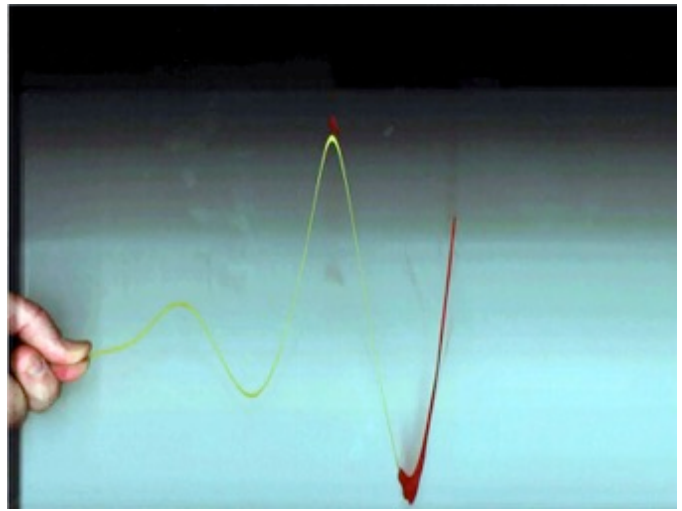

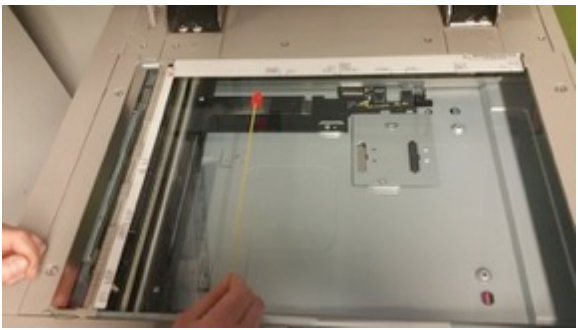

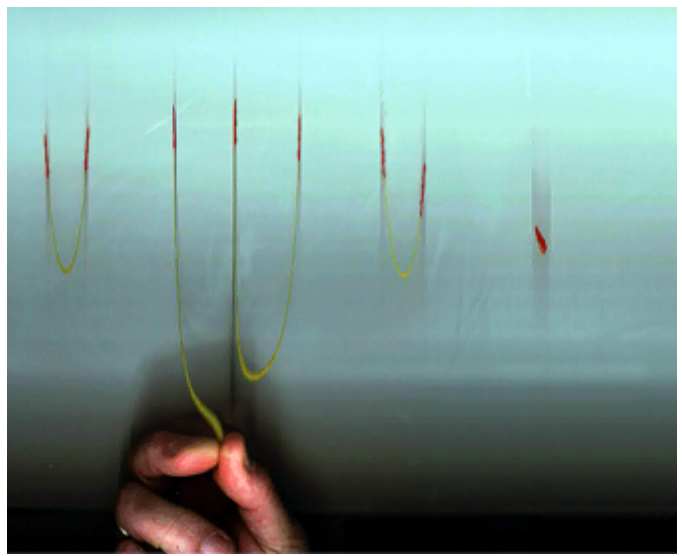

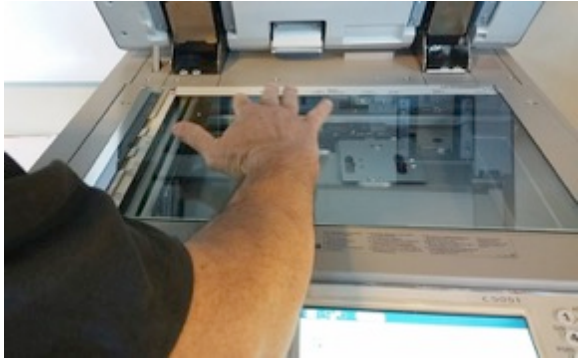

**Waving a hand in different directions**

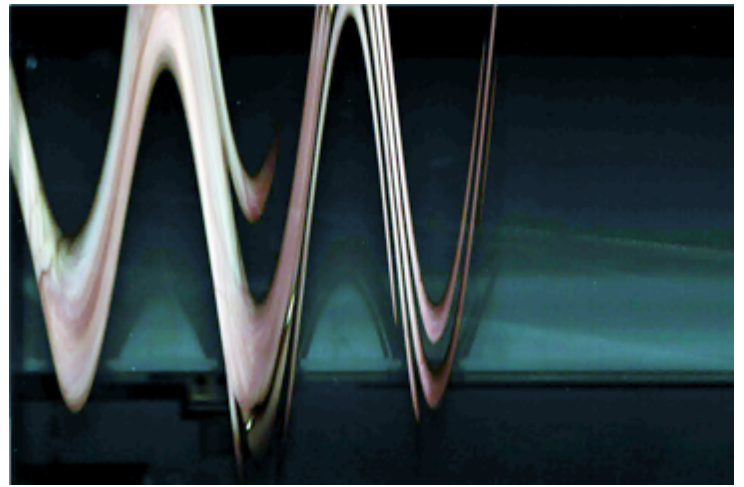

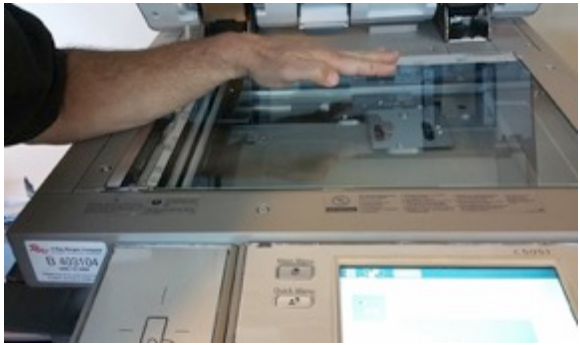

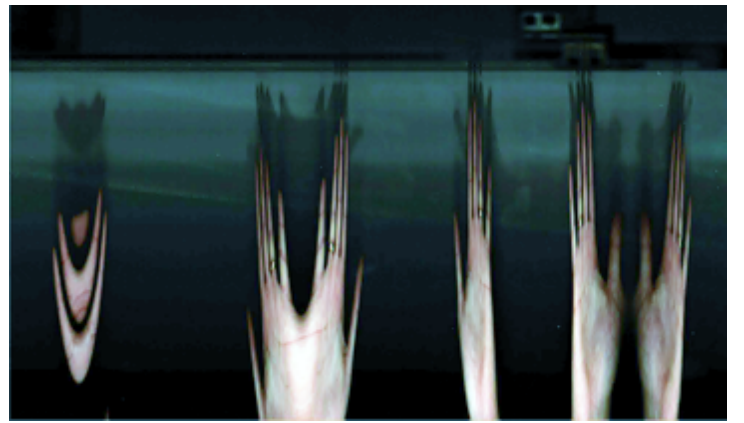

**Teacher Tips:**
**Here are some ideas to make your activity more palatable to others that don't share your interest in physics and art:**

- **Make sure to clean the glass scan bed after use.**
- **Printing your scans in black and white is usually cheaper than color.**
- **Scans sent digitally to your computer are cheaper and less resource intensive than printing (you will save on ink costs).**

So, don't ignore tension in your physics class, cut it with a knife and see what happens to the frequency!
And always remember to have fun, learn something and promote copying in science!

# A Musical Sidebar:
**How to tune a guitar with a photocopier!**
1. Guitar placed upside down on a scanner bed. Don't let the strings touch the glass.
2. Guitar strings should be strummed once as the photocopier scan bar passes underneath. Strumming an area not on the glass scan bed, like the over-hanging neck might lead to a clearer image.
    a. Strum all strings
    b. Pluck individual strings
    c. Strum a chord

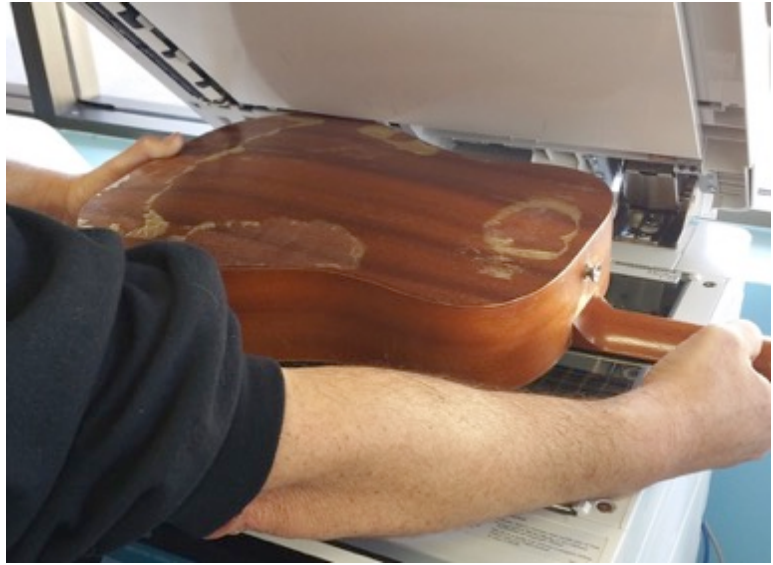

3. Check out your copy…..cool!

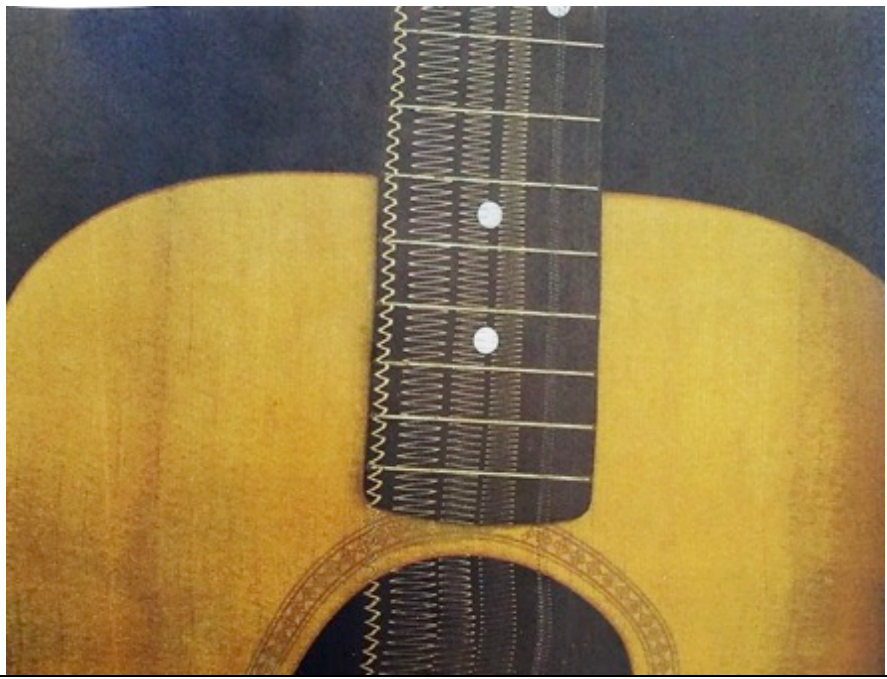

4. Measure your waves, just like your rubber band example above.
5. Calculate each string's frequency using the wave equation.
6. Compare your results from a guitar tuner or audio analysis software.

Close-up of strings on guitar neck.

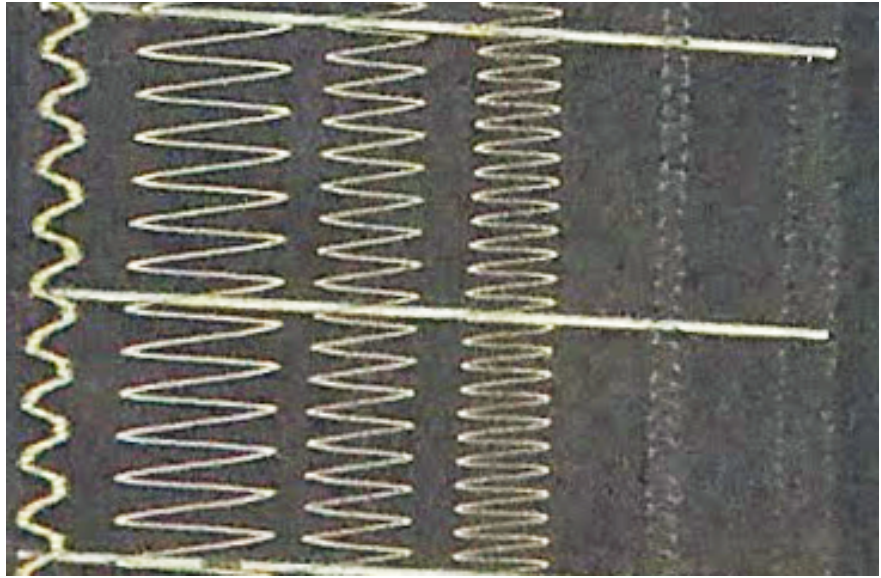

Here's how we did:

| Frequency | String 1 | String 2 | String 3 | String 4 | String 5 | String 6 |
|---|---|---|---|---|---|---|
| **Photocopier result** | 79hz | 109hz | 143hz | 193hz | 244hz | 315hz |
| **Audio tuner result** | 79hz | 107hz | 141hz | 189hz | 242hz | 320hz |
| **Standard tuning** | 82.4hz | 110hz | 147hz | 196hz | 247hz | 330hz |
| **Name of Note** | E | A | D | G | B | E |

We need to tune our guitar, but we won't fret about it!

Other things to notice:
- Which string was strummed first?
- Can you see overtones in the stings?
- Are there differences between nylon and metal strings?

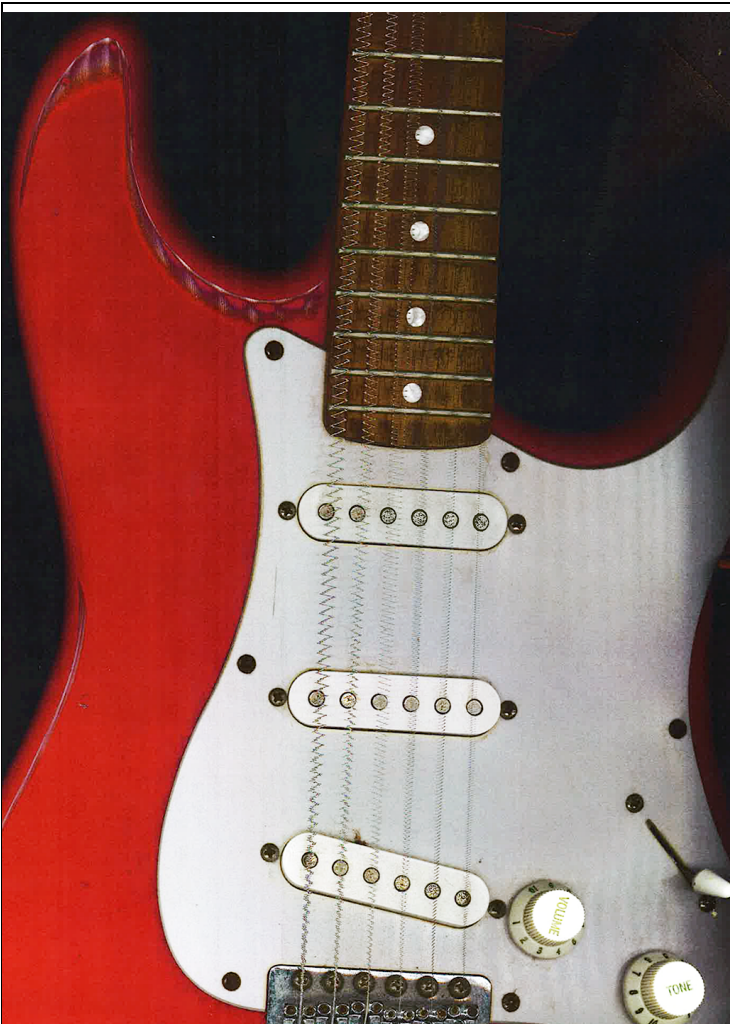
Electric Guitar

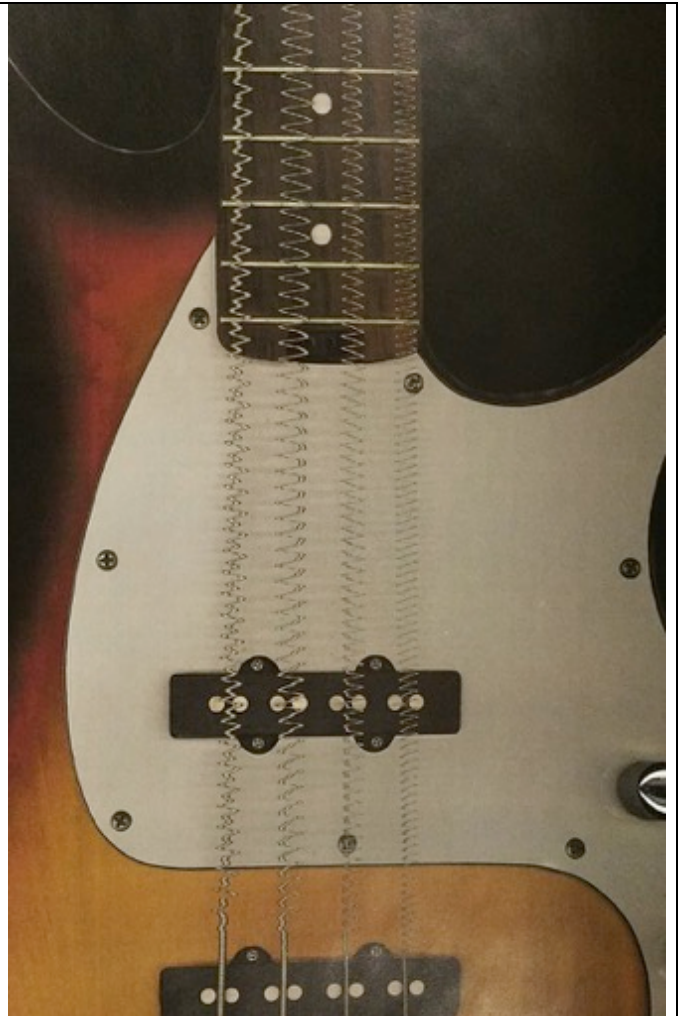
4-String Electric Bass

Acknowledgements:
Thanks for the help and support from The Exploratorium especially, Don Rathjen
(I would also like to thank: Zeke Kossover and the rest of the Exploratorium
Exploratorium Teacher Institute crew).

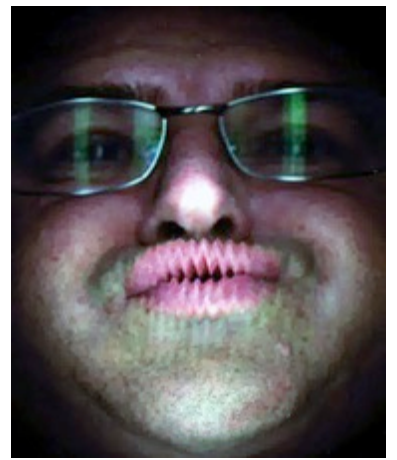

About the author:
Eric Muller, seen here, "blowing a raspberry" of 54Hz,
is a Sr. Science Educator at the Exploratorium Teacher Institute
in San Francisco.

References and online resources:

[1]Parker, Matt (2017, Jul 12)
"Rolling Shutter Explained on the Cheap"
*YouTube channel - standupmaths*
Retrieved from: https://www.youtube.com/watch?v=nP1elMR5qjc

[2]Sandlin, Destin (2017, Jun 30)
"Why Do Cameras Do This? (Rolling Shutter Explained) - Smarter Every Day 172"
*YouTube channel - SmarterEveryDay*
Retrieved from: https://www.youtube.com/watch?v=dNVtMmLlnoE

[3]MacIsaac, Dan
"Smartphones in a guitar redux"
*The Physics Teacher 53, 190 (2015);*
Retrieved from: https://doi.org/10.1119/1.4908097

[4]Tirosh, Udi (2012, September 24)
"Everything you wanted to know about rolling shutter"
*DIY Photography*
Retrieved from: https://www.diyphotography.net/everything-you-wanted-to-know-about-rolling-shutter/

[5]Muller, Eric (2016, September 9))
"Falling for Gravity"
*Exploratorium Science Snacks*
Retrieved from: https://www.exploratorium.edu/snacks/falling-gravity

---

This article will also we accessible via the "Physics Teacher" website at:
https://aapt.scitation.org/

As well as via Eric Muller's website at:
www.exo.net/~emuller/copierscience